\documentclass[12pt]{article}
\usepackage{epsf}
\def\la{\mathrel{\mathpalette\fun <}}
\def\ga{\mathrel{\mathpalette\fun >}}
\def\fun#1#2{\lower3.6pt\vbox{\baselineskip0pt\lineskip.9pt
\ialign{$\mathsurround=0pt#1\hfil
##\hfil$\crcr#2\crcr\sim\crcr}}}
\newcommand{\lan}{\left\langle}
\newcommand{\ran}{\right\rangle}

\newcommand{\veq}{\mbox{\boldmath${\rm q}$}}

\newcommand{\be}{\begin{equation}}
\newcommand{\ee}{\end{equation}}
\newcommand{\bc}{\begin{center}}
\newcommand{\ec}{\end{center}}

\newcommand{\mr}[1]{\mathrm{#1}}
\newcommand{\fr}[2]{\frac{#1}{#2}}
\newcommand{\lt}{\left}
\newcommand{\rt}{\right}

\newcommand{\rf}[1]{(\ref{#1})}
\newcommand{\lb}{\label}

\title{\bf A short distance quark-antiquark potential} 
\author{A.M. Badalian and D.S. Kuzmenko}
\date{\it Institute of Theoretical and Experimental
Physics,\\ 117218, B.Cheremushkinskaya 25, Moscow, Russia}

\begin{document}
\maketitle

\begin{abstract}
Leading terms of the static quark-antiquark potential in 
the background perturbation theory are reviewed, including 
perturbative, nonperturbative and interference ones. 
The potential is shown to describe lattice data at short
quark-antiquark separations with a good accuracy. 
\end{abstract}

1. The static quark-antiquark potential was calculated
with high accuracy in lattice QCD some years ago \cite{1}.
It was shown to be well described by the phenomenological
Coulomb+linear Cornell potential at sufficiently large 
quark-antiquark separations, $R\ga 0.2$ fm. At smaller distances 
the Cornell potential is not applicable. The region 0.03 fm $\leq 
R \leq$ 0.22 fm was studied in quenched lattice theory in detail 
\cite{2}, and the conclusion was made that the standard 
perturbative theory expansion in coupling constant does not yield 
appropriate description of lattice results, at least in one- and 
two-loop approximations. 
As is known, next terms of the asymptotic 
coupling expansion depend on the renormalization scheme, and so 
the corresponding static potential does.
One can argue that the standard perturbative theory fails because 
this region is close to the unphysical Landau pole  of the 
strong coupling.

There is a wealth of literature on the short distance potential 
behavior, see \cite{8}, \cite{Lit1}, \cite{4}, \cite{Lit2} and 
references therein. In the talk we consider the static 
quark-antiquark potential in the background perturbation theory 
(BPT) \cite{3}. This potential incorporates both the features of 
the standard perturbative potential at tiny distances, $R\la 0.05$ 
fm, and of the Cornell potential at $R\ga 0.4$ fm due to taking 
nonperturbative background field effects into account. 
After  brief review of leading background potential 
terms we present our results concerning the 
behavior of the potential at short distances and its comparison 
with the lattice \cite{4}. 

2. The gluon field $A_\mu$ in BPT 
is divided into the dynamical perturbative part $a_\mu$
and  the background nonperturbative field $B_\mu$,
\be
A_\mu=a_\mu+B_\mu.
\lb{1}
\ee 
The background field, in which perturbative valence gluons 
propagate, results in the vacuum condensate creation. 

The static potential has to be calculated using the vacuum 
averaged Wilson loop for the quark-antiquark pair. The BPT Wilson 
loop expansion in the field $a_\mu$ has the form \cite{3}
\be
W(B+a)=W(B)+\sum_{n=1}^\infty(ig)^n W^{(n)}(B;x(1)..x(n))a_{\mu_1}...
a_{\mu_n} dx_{\mu_1}(1)...dx_{\mu_n}(n).
\lb{2}
\ee
To perform the averaging of the expression \rf{2}, we take into 
account that the linear in $a_\mu$ term vanishes, 
\be
\lan W(B+a)\ran_{B,a}=\lan W(B)\ran_B-g^2\lan W^{(2)}(B;x,y)\ran_B
dx\, dy+...,
\lb{3}
\ee
where 
\be
-g^2 W^{(2)}dx\, dy =-g^2\int
\Phi^{\alpha\beta}(x,y,B)t^a_{\delta\alpha}t^b_{\beta\gamma}
G^{ab}_{\mu\nu}(x,y,B)\Phi^{\gamma\delta}(y,x) dx_\mu\, dy_\nu.
\lb{4}
\ee
The Green function of the valence gluon in the background gauge
takes the form \cite{3}
\be
G_{\mu\nu}(x,y)=\lan a_\mu(x)\,a_\nu(y)\ran_B=
\lan x|(\hat D^2_\lambda \cdot \delta_{\mu\nu}-2ig\hat
F_{\mu\nu})^{-1} |y\ran,
\lb{5}
\ee
where $\hat D^2_\lambda$ is the covariant derivative depending on 
the field $B$, and $\hat F_{\mu\nu}$ is the background field 
strength. The operator  $\hat F_{\mu\nu}$ has to be considered
as a correction \cite{3}. The Green function expansion
in the $\hat F_{\mu\nu}$ takes the form
\be
G(x,y)
=\lan x|D^{-2}|y\ran-\lan x|D^{-2} 2ig \hat F\, D^{-2}|y\ran+
\lan x|D^{-2} 2ig \hat F\, D^{-2} 2ig \hat F\, D^{-2}|y\ran+... .
\lb{6}
\ee
The terms of odd powers in the field $B$ vanish. Let us confine 
ourselves by the third term in the expansion \rf{6} and quadratic 
term in \rf{3}, and write the Wilson loop in the form
\be
\lan W(B+a)\ran_{B,a}\simeq\lan W(B)\ran_B+\lan W(B)\ran_B
\lt(\tilde W^{(2)}_1+\tilde W^{(2)}_3\rt).
\lb{7}
\ee
One can verify using  the Fock-Feynman-Shwinger
representation \cite{6} for the Green function expansion
\rf{6} that the factorization of the second term in \rf{7} 
is valid. Terms proportional to $\tilde W^{(2)}_1$
and $\tilde W^{(2)}_3$ come from the first and third terms
in \rf{6}. The following approximate expression is valid
within the accuracy considered,
\be
\lan W(B+a)\ran_{B,a}\simeq \lan W(B)\ran_B\exp\lt(\tilde 
W^{(2)}_1+\tilde W^{(2)}_3\rt).
\lb{8}
\ee
Taking the logarithm, we arrive at the three corresponding terms
in the static potential, 
\be
  V_{Q\bar Q}(r) =- \lim_{T\to \infty} \frac{1}{T}\ln \langle
 W(B+a)\rangle_{B,a}=V_{NP}(r)+V_P(r)+V_{\mr{int}}(r),
\lb{9}
\ee
where $r$ is the quark-antiquark separation,
the nonperturbative potential $V_{NP}$ is given by the 
average of the Wilson loop $\lan W(B)\ran_B$, the background 
perturbative potential $V_P$ comes from the first term in
the expansion \rf{6}, and the interference potential 
$V_{\mr{int}}$ comes from the third one. The nonperturbative 
potential rises linearly at distances $r\ga T_g$, where 
$T_g=0.12\div 0.2$ fm is the background field correlation length, 
and is quadratic at short distances $r\la T_g$ (see e.g. the talk 
at this conference \cite{7}), 
\be
V_{NP}(r)\simeq \frac2\pi \frac{r}{T_g}\, \sigma r.
\lb{10}
\ee
One can see that at short distances $V_{NP}(r)\ll \sigma r$.

The interference potential was calculated in \cite{8} and shown 
to be close to the linear one at short distances,
\be
 V_{\mr{int}}(r)\simeq\sigma R.
\lb{11}
\ee

The background perturbative potential has the form \cite{3}
\be
V_P(r)=-\frac{C_F\alpha_B(r)}{r},
\lb{12}
\ee
where  $C_F=4/3$ and the background coupling $\alpha_B(r)$ 
saturates with some critical, or freezing, value at large $r$. 

3. We proceed now to a  comparison between background and 
standard couplings at short distances. The Callan-Symanzik 
equation yields the following expressions for the running 
coupling constant in one- and two-loop  approximations,

\be
\alpha_s^{(1)}(q)=\frac{4\pi}{\beta_0 \ln\frac{q^2}{\Lambda^2}},
\lb{13}
\ee

\be
\alpha_s^{(2)}(q)=\alpha_s^{(1)}(q)
\left(1-\frac{\beta_1}{\beta_0^2}\frac{\ln
\ln\frac{q^2}{\Lambda^2}}{\ln\frac{q^2}{\Lambda^2}}\right),
\lb{14}
\ee
where $\beta_0=11-\frac2 3 n_f$, $\beta_1=102-\frac{38}{3} n_f$,
$q^2\equiv \veq^2$ and $\Lambda\approx 385$ MeV is the QCD 
constant (for the discussion of its value see \cite{4}). 

The modified Callan-Symanzik equation is used in BPT \cite{3}
for the background coupling $\alpha_B$,
which takes into account the background field contribution and 
leads to the substitution $q^2\to q^2+m_B^2$ in \rf{13}, \rf{14},
where $m_B=1$ GeV \cite{4}. 
One can see that the background coupling saturates with the 
freezing value in infrared region $q^2\ll m_B^2$  and turns
to standard $\alpha_s$ in ultraviolet one.

The background coupling in the coordinate representation
can be calculated using the Fourier transform, and in two-loop
approximation takes the form

\begin{figure}[!t]
  \epsfxsize=12.5cm
  \hspace*{2.1cm}   
\epsfbox{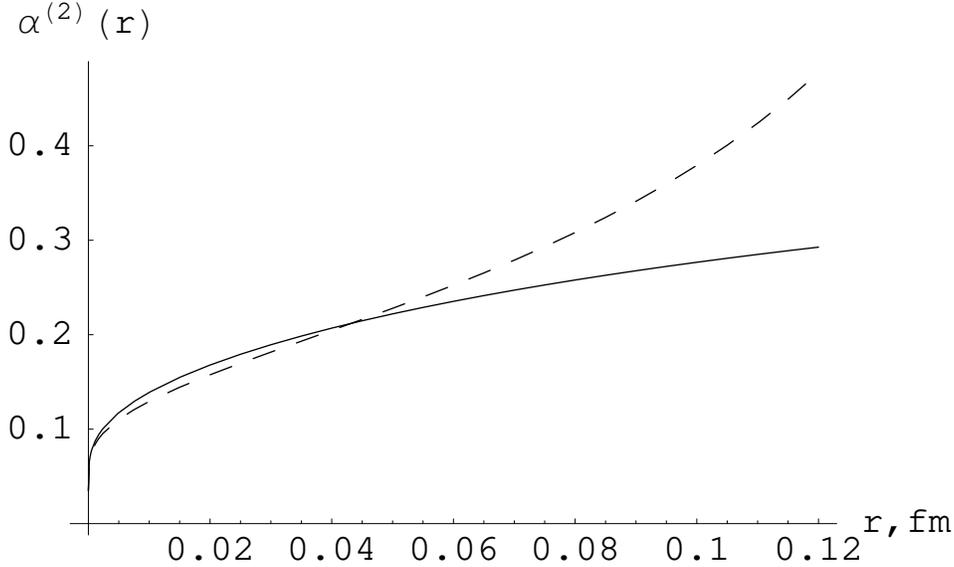}
\caption{The  background coupling $\alpha_B^{(2)}(r)$ with
$\Lambda=385$ MeV, $m_B=1.0$ GeV (solid line) compared to the
perturbative  $\alpha_s^{(2)}(r)$ with 
$\Lambda_R=686$ MeV (dashed line) at short distances.
A rigorous definition of $\alpha_s(r)$ and $\Lambda_R$ is given 
in \cite{4}.}   
\end{figure}

\be
 \alpha_B^{(2)}(r)=\frac8{\beta_0}\int^\infty_0 \fr{dq}q
\frac{\sin qr}{\ln\frac{q^2+m_B^2}{\Lambda^2}}
\left(1-\frac{\beta_1}{\beta_0^2}\frac{\ln
\ln\frac{q^2+m_B^2}{\Lambda^2}}{\ln\frac{q^2+m_B^2}{\Lambda^2}}
\right).
\lb{15}
\ee

It is shown in comparison with the standard coupling in Fig. 1.
One can see from the figure considerable difference 
between two curves at $r\ga 0.05$ fm. The coupling 
$\alpha_s(r)$ grows rapidly in this region due to the influence of 
the pole, which is situated at  $r\simeq 0.3$ fm. 

Let us compare now the background static potential
behavior at short distances with the
lattice one. 

Relying on the relations \rf{9}-\rf{12}, \rf{15}, we approximate
the potential in this region by the sum \cite{4}
\be
V_B(r)\approx -\frac43 \fr{\alpha_B^{(2)}(r)}{r}+\sigma r.
\lb{16}
\ee
The behavior of $V_B(r)$ at $r<0.22$ fm is shown in Fig. 2
in comparison with lattice points from \cite{2}.
The values of $\sigma=0.2$ GeV$^2$ and overall shift $C=-253$ MeV
were taken from the fit, which has provided agreement between
background and lattice potentials with the accuracy $\la 50$ MeV
of the latter.

4. In summary we enumerate some properties of the 
 short distance static  quark-antiquark potential.
\begin{itemize}
\item
The potential at $r\ll T_g$ consists mainly of perturbative
and interference parts. Purely nonperturbative potential
is small in this region.
\item
The background running coupling constant saturates with the 
freezing value in infrared region and goes over to the standard
coupling in ultraviolet region.
\item
A considerable difference between standard and background 
couplings in two-loop approximation starts already at 
distances $r\ga 0.05$ fm. 
\item
The background potential, approximated as a sum of two-loop
background perturbative potential and linear  potential with the 
slope $\sigma$, yields a good description of lattice 
simulations at short distances. This in turn means that
the short distance area law for the Wilson 
loop, used  in particular in the QCD string approach
\cite{5}, is justified.
\end{itemize}

This work has been supported by 
 RFBR grants 00-02-17836,  00-15-96786, and INTAS 00-00110, 
00-00366.

\newpage
\bc
{\bf \Large Figure caption}
\ec 

Figure 2:   The background  potential $V_B(r)$  (solid line) 
compared to the lattice one from \cite{2} (points) in units of 
$r_0=2.5$ GeV$^{-1}$. 1-loop and 2-loop standard 
perturbative potentials from  \cite{2} are shown by thin solid 
line below lattice points and dash-dotted line correspondingly; 
1-loop + linear with the large slope $\sigma^*\approx 1$ GeV$^2$ 
potential is shown by dashed line. The figure is taken from
our paper \cite{4}.

\end{document}